\documentclass[12pt]{article}
\usepackage{graphicx}

\newcommand\pubnumber{DPF2015-261}
\newcommand\pubdate{\today}

\def\pitt{Department of Physics and Astronomy\\
University of Pittsburgh, Pittsburgh, Pennsylvania 15260, USA}

\def\Title#1{\begin{center} {\Large #1 } \end{center}}
\def\Author#1{\begin{center}{ \sc #1} \end{center}}
\def\Address#1{\begin{center}{ \it #1} \end{center}}

\newcommand\pubblock{\rightline{\begin{tabular}{l} \pubnumber\\
         \pubdate  \end{tabular}}}
\newenvironment{Abstract}{\begin{quotation}  }{\end{quotation}}
\newenvironment{Presented}{\begin{quotation} \begin{center} 
             PRESENTED AT\end{center}\bigskip 
      \begin{center}\begin{large}}{\end{large}\end{center} \end{quotation}}

\def\beq{\begin{equation}}
\def\eeq#1{\label{#1}\end{equation}}
\def\eeqn{\end{equation}}

\def\beqa{\begin{eqnarray}}
\def\eeqa#1{\label{#1}\end{eqnarray}}
\def\eeqan{\end{eqnarray}}

\let\bar=\overbar

\def\Dslash{\not{\hbox{\kern-4pt $D$}}}
\def\dslash{\not{\hbox{\kern-2pt $\del$}}}

\def\msb{{\bar{\ssstyle M \kern -1pt S}}}

\begin{document}
\begin{titlepage}
\pubblock

\vfill
\Title{Search for Flavor Changing Non-standard Interactions with the MINOS+ Experiment}
\vfill
\Author{ Nick Graf}
\Address{\pitt}
\vfill
\begin{Abstract}
A study of MINOS+ sensitivity to non-standard interactions and previously published results using MINOS data are presented.
\end{Abstract}
\vfill
\begin{Presented}
DPF 2015\\
The Meeting of the American Physical Society\\
Division of Particles and Fields\\
Ann Arbor, Michigan, August 4--8, 2015\\
\end{Presented}
\vfill
\end{titlepage}

\section{Introduction}

MINOS+ began operation in September 2013 as an extension of the MINOS experiment \cite{minosprd}, and extends the study of neutrino oscillations as well as the search for exotic physics such as flavor changing non-standard interactions (NSI) over a baseline of 735 km.   The experiment uses two magnetized tracking calorimeters placed in the NuMI beam
at Fermilab.  To date MINOS+ has reported the most precise measurement of neutrino oscillations in the atmospheric sector from a three-neutrino flavor analysis of the full $14.3 \times 10^{20}$ PoT MINOS era beam sample combined with the full MINOS and MINOS+ atmospheric neutrino samples.

MINOS/MINOS+ is able to separately produce well understood beams enriched in neutrinos and antineutrinos and to identify neutrinos and antineutrinos on an event-by-event basis in its magnetized detectors.  This allows sensitive searches for NSI in the $\mu$-$\tau$ and e-$\tau$ sectors.  We present results using a partial MINOS beam data set and projected sensitivity for the full MINOS and MINOS+ beam data set.

\section{Neutrino Oscillations and Non-standard Interactions in MINOS}

It is well established from experiment \cite{minosprd, superk, dayabay} that neutrinos undergo flavor change as the propagate, which is explained but the quantum mechanical mixing of neutrino flavor and mass eigenstates.  This mixing is parameterized by three angles, $\theta_{12}$, $\theta_{13}$, $\theta_{23}$, and a CP-violating phase, $\delta$ \cite{pont}.  The presence of matter allows for alternative flavor changing mechanisms, such as the Mikheyev-Smirnov-Wolfenstein (MSW) effect \cite{msw}, which alters the survival probability of electrons neutrinos only, through interactions with electrons in the surrounding medium.

The survival probability of muon and tau neutrinos could be alternated by NSI \cite{nsi} with matter in a similar way to standard matter effects.  The NSI Hamiltonian can be included as a perturbation to standard oscillations.  In the two flavor approximation:

\begin{equation}
H_{NSI} = V\left( \begin{array}{cc}
\epsilon_{\mu\mu} & \epsilon_{\mu\tau} \\
\epsilon^{*}_{\mu\tau} & \epsilon_{\tau\tau} \end{array} \right),
\label{eq:ham}
\end{equation} 

\noindent  where the coefficients $\epsilon_{\alpha\beta}$ give the strength of the NSI effect on transitions between flavors $\alpha$ and $\beta$.  Considering only flavor changing NSI, the survival probability can be written as

\begin{equation}
P \left(\nu_{\mu} \rightarrow \nu_{\mu} \right) = 1 - \left[1-cos^{2}(2\theta)\frac{L^{2}_{m}}{L^{2}_{0}} \right]sin^{2} \left( \frac{L}{L_{m}} \right),
\label{eq:survprob}
\end{equation} 

\noindent  where

\begin{equation}
L_{m} = \frac{L_{0}}{[1 \pm 2sin(2\theta)L_{0}\epsilon_{\mu\tau}|V| + (L_{0}\epsilon_{\mu\tau}|V|)^{2}]^{\frac{1}{2}}},
\label{eq:osclen}
\end{equation} 

\noindent  $L_{0} = \left(\frac{4E}{\Delta m^{2}} \right)$, $\Delta m^{2} = \Delta m^{2}_{\alpha\beta} = m^{2}_{\alpha} - m^{2}_{\beta}$, $E$ is the neutrino energy, $\theta$ is the mixing angle, and $L$ is the neutrino path length.

Accelerator-based oscillation experiments, and MINOS+ in particular, provide a powerful tool for NSI searches with the ability to produce well understood beams of neutrinos and antineutrinos separately, as NSI affects them differently.  Since the MINOS detectors are magnetized, they have the capability to identify neutrinos and antineutrinos on an event-by-event basis.  While short-baseline neutrino experiments have constrained NSI \cite{nutev}, the sensitivity is improved in long-baseline experiments from using the Earth's matter along the neutrino path as the interaction medium.  For these reasons MINOS+ is well-suited to constrain NSI in the $\mu$-$\tau$ mixing sector.

\section{Results}

The results presented here are based on an exposure of $7.09 \times 10^{20}$ POT in neutrino mode, combined with a $2.95 \times 10^{20}$ POT exposure in antineutrino mode \cite{prd_nsi}.  Due to the opposite sign of the matter potential in Eq.~\ref{eq:survprob} for neutrinos and antineutrinos, NSI, if present, will alter the survival probability of neutrinos and antineutrinos in opposite directions. The magnitude of $\epsilon_{\mu\tau}$ is proportional to the difference in probability between neutrinos and antineutrinos, and the sign of $\epsilon_{\mu\tau}$ is proportional to the difference in is determined by the sign of the probability difference.

The FD neutrino and antineutrino spectra in the absence of flavor change are predicted using the ND data by first correcting the ND spectra for inefficiency and backgrounds and then extrapolating to the FD by a transfer matrix obtained from simulation \cite{minosprd}.
They are fit simultaneously to three parameters, $|\Delta m^{2}|$, $sin^{2}(2\theta)$, and $\epsilon_{\mu\tau}$, in the combined oscillation and NSI model in Eq.~\ref{eq:survprob}, using a binned log-likelihood. The value of the mixing angle is constrained to be physical by asserting $0 \leq sin^{2}(2\theta) \leq 1$.

The overall systematic uncertainty in the measurement is much smaller than the statistical uncertainty.  The four most significant sources are: i) the hadronic energy scale, ii) the muon energy scale, iii) the NC background, and iv) the relative normalization between the Near and Far detectors.  The are included in the fit using penalty terms. The best fit parameters from this procedure are found to be

\begin{center}
$\Delta m^{2} = 2.39^{+0.14}_{-0.11} \times 10^{-3}$ eV$^{2}$,

$sin^{2}(2\theta) = 1.00^{+0.00}_{-0.06}$,

$\epsilon_{\mu\tau} = -0.07^{+0.08}_{-0.08}$,
\end{center} 

\noindent  with the allowed region $-0.20 < \epsilon_{\mu\tau} < 0.07$ (90\% C.L.).

The allowed regions of fit parameters are shown in Fig.~\ref{fig:nsi_sens}, where three two-dimensional slices from a 3D likelihood surface are chosen by marginalizing over the third parameter.  Within errors the fit is consistent with no contribution to flavor change from NSI and is in good agreement with previously published results \cite{superk2, minos_nsi}, as well as with values of $\epsilon_{\mu\tau}$ extracted from global fits to data from multiple experiments \cite{blennow1, biggio, escrihuela}.

\begin{figure}[htb]
\centering
\includegraphics[height=1.4in]{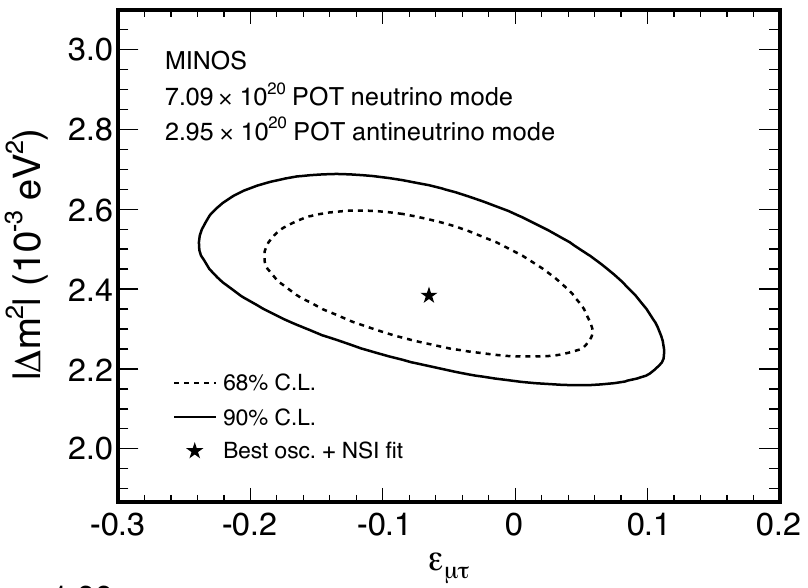}
\includegraphics[height=1.4in]{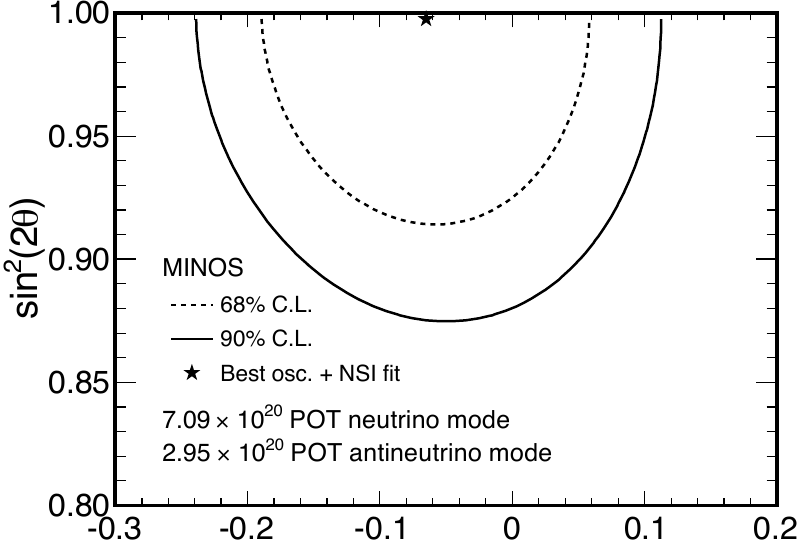}
\includegraphics[height=1.4in]{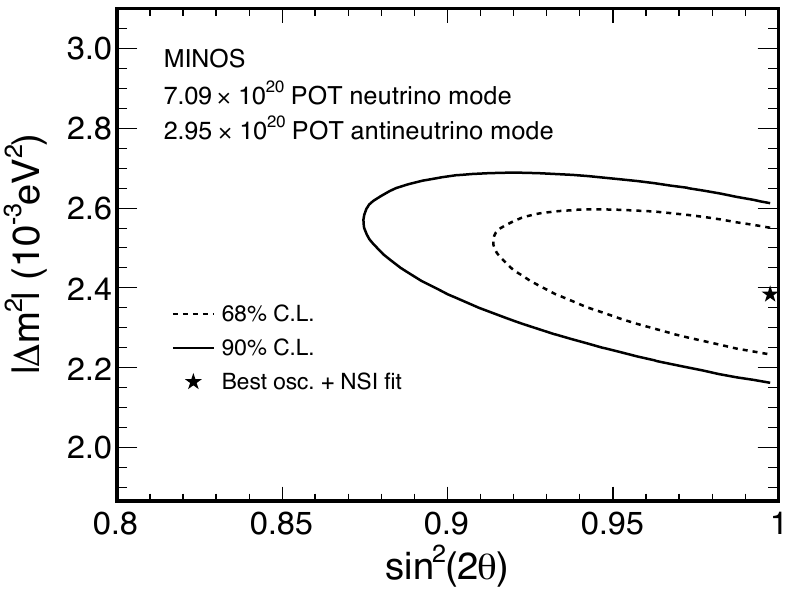}
\caption{68\% and 90\% C.L. intervals from an NSI search using MINOS data \cite{prd_nsi}.  A total of $7.09 \times 10^{20}$ POT in neutrino mode and $2.95 \times 10^{20}$ POT in antineutrino mode were analyzed.}
\label{fig:nsi_sens}
\end{figure}

\section{Projected Sensitivity}

Sensitivity to NSI in the $\mu$-$\tau$ mixing sector can be improved by including additional MINOS/MINOS+ data.  The black curves in Fig.~\ref{fig:nsi_sens} shows projected sensitivity contours when including the complete MINOS beam data set.   Additional projections are shown that include MINOS+ data: $3 \times 10^{20}$ POT $\nu_{\mu}$-mode (green), $10 \times 10^{20}$ POT $\nu_{\mu}$-mode (blue), and $10 \times 10^{20}$ POT $\nu_{\mu}$-mode plus $4 \times 10^{20}$ POT $\bar{\nu}_{\mu}$-mode (red).  We expect an improvement of 17\% from inclusion of the first year of MINOS+ data.  An improvement of 50\% is projected for inclusion of the full MINOS+ data set, with nearly half of the improvement contributed by the antineutrino data set.
\begin{figure}[htb]
\centering
\includegraphics[height=1.5in]{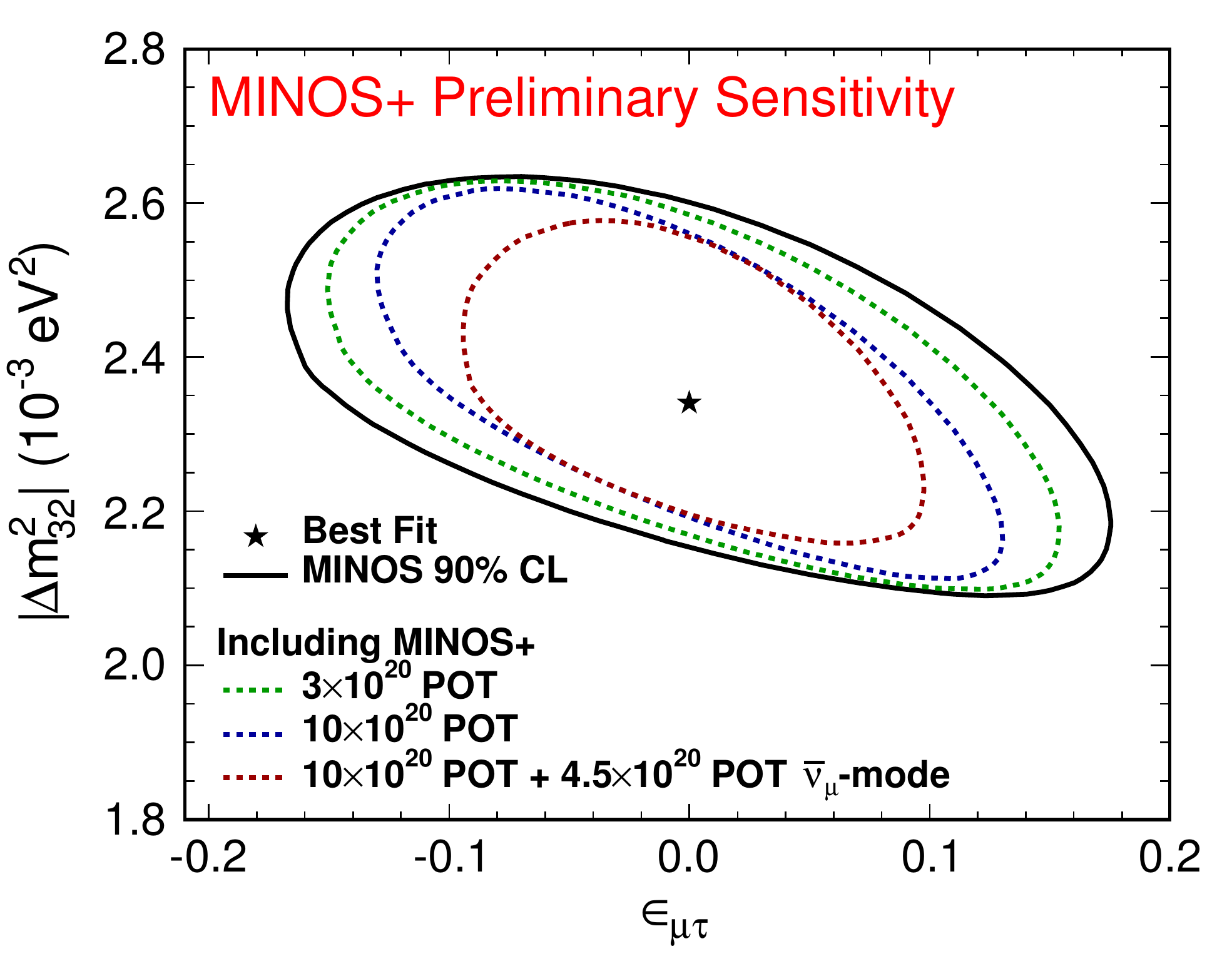}
\includegraphics[height=1.5in]{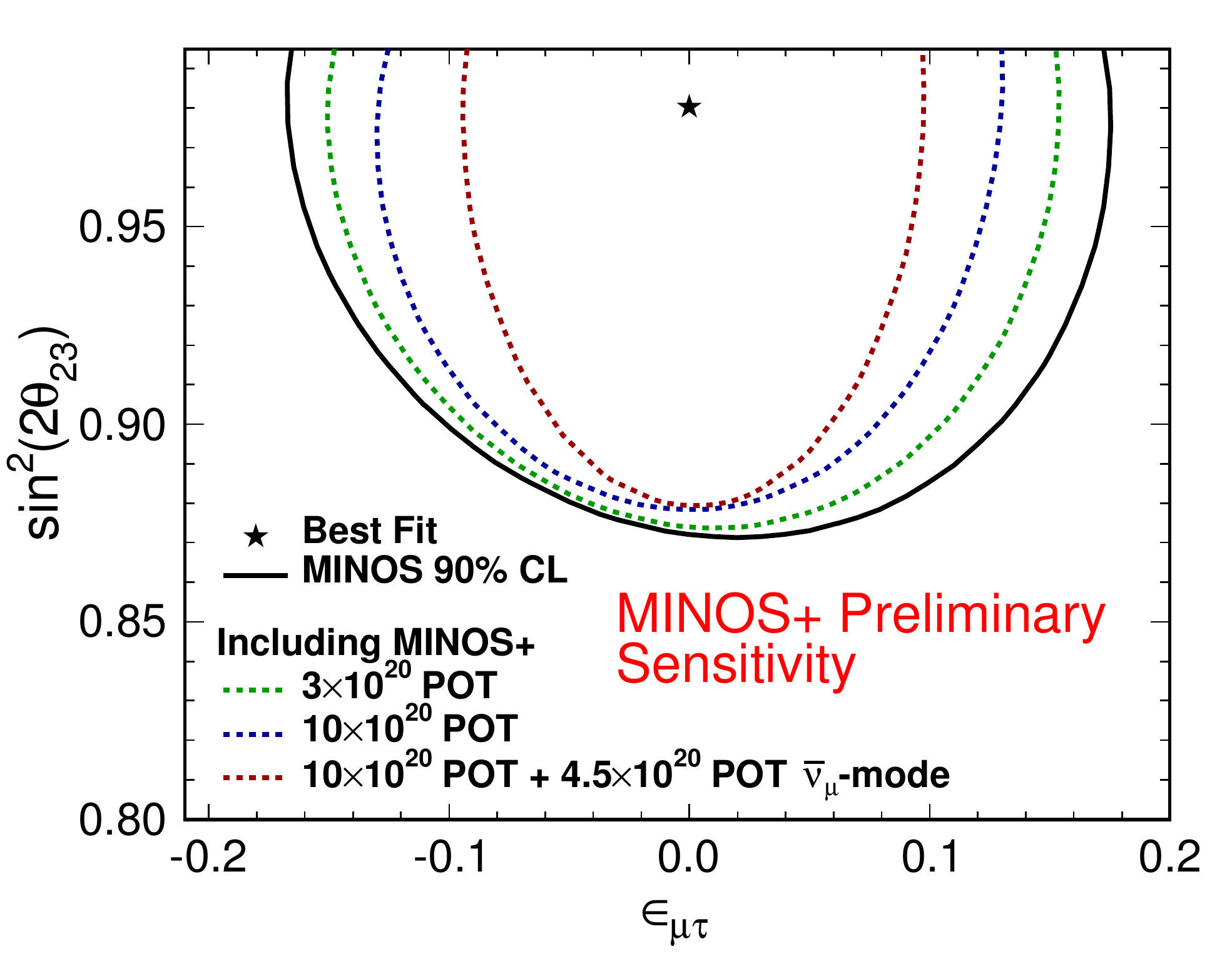}
\includegraphics[height=1.5in]{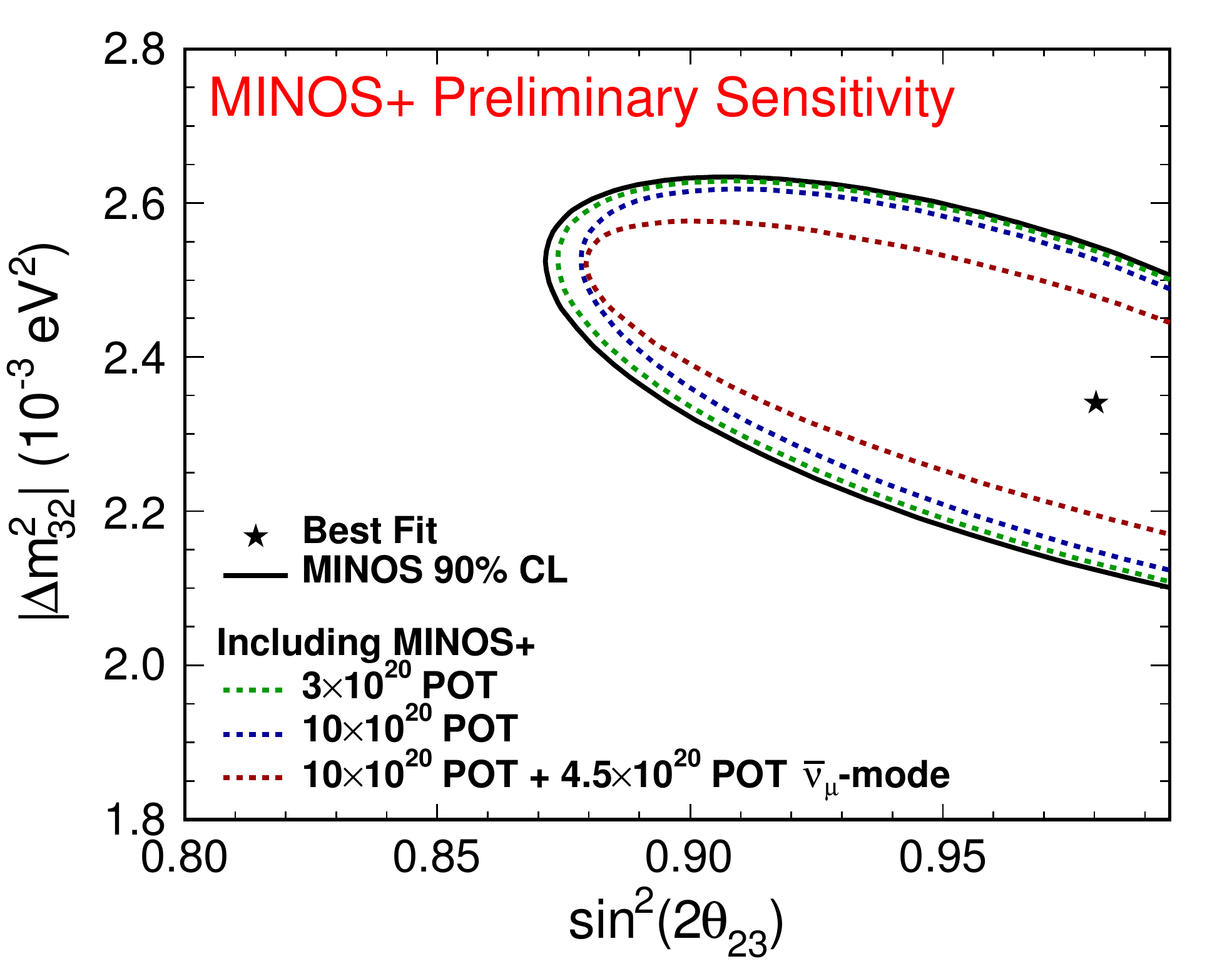}
\caption{Projected sensitivity to NSI in the $\mu$-$\tau$ mixing sector after including the full MINOS low energy beam data set and additional MINOS+ data.}
\label{fig:nsi_sens}
\end{figure}

\end{document}